\documentclass[useAMS,usenatbib]{mnras}

\usepackage{ulem}
\usepackage{color}
\usepackage{graphics,graphicx}
\usepackage{times} 
\usepackage{amssymb}
\usepackage{amsmath}
\usepackage{natbib}
\usepackage{url}
\usepackage{multirow}
\usepackage{ctable}
\usepackage{threeparttable}
\newif\ifAMStwofonts
\AMStwofontstrue

\def\xmm{{\it XMM-Newton}}

\def\suzaku{{\it Suzaku}}

\def\epicpn{{EPIC-pn}}
\def\epicmos1{{EPIC-MOS1}}
\def\epicmos2{{EPIC-MOS2}}
\def\epicmos{{EPIC-MOS}}

\def\nustar{{\it NuSTAR}}

\def\deg{$^{\circ}$}

\def\kmps{\hbox{$\rm\thinspace km~s^{-1}$}}
\def\pcmsq{\hbox{$\rm\thinspace cm^{-2}$}}
\def\H0{{\rm km~s$^{-1}$~Mpc$^{-1}$}}

\def\kev{\hbox{\rm keV}}

\def\ctps{\hbox{$\rm\thinspace ct~s^{-1}$}}

\def\ergpcmsqps{\hbox{$\rm\thinspace erg~cm^{-2}~s^{-1}$}}

\def\ergps{\hbox{erg~s$^{-1}$}}
\def\ergcmps{\hbox{\rm erg~cm~s$^{-1}$}}

\def\msun{\hbox{$\rm M_{\odot}$}}

\def\kte{$kT_{\rm{e}}$}

\def\chisq{{$\chi^{2}$}}

\def\xspecv{\hbox{\small XSPEC}\, v12.6.0f}

\def\nustardas{{\small NUSTARDAS}}

\def\xmmselect{\hbox{\rm{\small XMMSELECT}}}

\def\addascaspec{\hbox{\rm{\small ADDASCASPEC~\/}}}
\def\flx2xsp{\rm{\small FLX2XSP}}
\def\sas{\hbox{\rm{\small SAS~\/}}}
\def\epchain{\hbox{\rm{\small EPCHAIN}}}
\def\emchain{\hbox{\rm{\small EMCHAIN}}}

\def\rmfgen{\hbox{\rm{\small RMFGEN}}}
\def\arfgen{\hbox{\rm{\small ARFGEN}}}

\def\addascaspec{\rm{\small ADDASCASPEC}}
\def\nupipeline{{\small NUPIPELINE}}
\def\nuproducts{{\small NUPRODUCTS}}

\def\xstar{\hbox{{\small XSTAR}}}
\def\grid25{\hbox{\rm{\small GRID25}}}

\def\tbabs{{\small TBABS}}

\def\diskbb{\rm{\small DISKBB}}

\def\xillver{{\small XILLVER}}
\def\xillvercp{{\small XILLVER\_CP}}

\def\gsmooth{{\small GSMOOTH}}

\def\relxill{\rm{\small RELXILL}}

\def\relxilllpcp{{\small RELXILLLP\_CP}}

\def\borus{{\small BORUS}}

\def\nthcomp{\rm{\small NTHCOMP}}

\def\cutoffpl{\rm{\small CUTOFFPL}}
\def\mekal{{\small MEKAL}}

\def\fexxv{\hbox{\rm Fe\,{\small XXV}}}
\def\fexxvi{\hbox{\rm Fe\,{\small XXVI}}}

\def\eg{{\it e.g.}}

\def\ie{{\it i.e.~\/}}

\def\la{\mathrel{\hbox{\rlap{\hbox{\lower4pt\hbox{$\sim$}}}{\raise2pt\hbox{$<$}}}}}
\def\ga{\mathrel{\hbox{\rlap{\hbox{\lower4pt\hbox{$\sim$}}}{\raise2pt\hbox{$>$}}}}}

\def\d25{D$_{25}$}
\def\nh{{$N_{\rm H}$}}

\def\.25{0.25 keV\thinspace}

\def\rg{$R_{\rm{G}}$}
\def\rh{$R_{\rm{H}}$}

\def\Rfrac{$R_{\rm{frac}}$}

\def\iras{IRAS\,00521--7054}
\def\nsims{10,000}
\def\ngtr{none}
\def\ngtrexcl{151}
\def\ngtrAl{none}

\title[A Low-Flux State in IRAS\,00521--7054]{A Low-Flux State in IRAS\,00521--7054 seen with \textit{NuSTAR} and \textit{XMM-Newton}: Relativistic Reflection and an Ultrafast Outflow}

\author[D.\,J. Walton et al.]
{\parbox{7.in}{D.\,J. Walton$^{1}$\thanks{E-mail: dwalton@ast.cam.ac.uk},
E. Nardini$^{2}$, 
L. C. Gallo$^{3}$, 
M. T. Reynolds$^{4}$, 
C. Ricci$^{5,6,7}$ 
T. Dauser$^{8}$, \\ 
A. C. Fabian$^{1}$, 
J. A. Garc\'ia$^{9,8}$,
F. A. Harrison$^{9}$,
G. Risaliti$^{10}$,
D. Stern$^{11}$ 
\\[0.2cm]
\footnotesize
$^{1}$ \it{Institute of Astronomy, University of Cambridge, Madingley Road, Cambridge CB3 0HA, UK} \\
$^{2}$ \it{INAF -- Osservatorio Astrofisico di Arcetri, Largo Enrico Fermi 5, I-50125 Firenze, Italy} \\
$^{3}$ \it{Department of Astronomy and Physics, Saint Mary's University, 923 Robie Street, Halifax, NS, B3H 3C3, Canada} \\
$^{4}$ \it{Department of Astronomy, University of Michigan, 1085 South University Ave, Ann Arbor, MI 48109-1107, USA} \\
$^{5}$ \it{Nucleo de Astronomia de la Facultad de Ingenieria, Universidad Diego Portales, Av. Ejercito Libertador 441, Santiago, Chile} \\
$^{6}$ \it{Kavli Institute for Astronomy and Astrophysics, Peking University, Beijing 100871, China} \\
$^{7}$ \it{Chinese Academy of Sciences South America Center for Astronomy, Camino El Observatorio 1515, Las Condes, Santiago, Chile} \\
$^{8}$ \it{Dr. Karl Remeis-Observatory and Erlangen Centre for Astroparticle Physics, Sternwartstr. 7, 96049 Bamberg, Germany} \\
$^{9}$ \it{Space Radiation Laboratory, California Institute of Technology, Pasadena, CA 91125, USA} \\
$^{10}$ \it{Dipartimento di Fisica e Astronomia, Universita di Firenze, via G. Sansone 1, 50019 Sesto Fiorentino, Firenze, Italy} \\
$^{11}$ \it{Jet Propulsion Laboratory, California Institute of Technology, Pasadena, CA 91109, USA} \\
}}
\date{}

\begin{document}
\pagerange{\pageref{firstpage}--\pageref{lastpage}}
\maketitle
\label{firstpage}

\begin{abstract}
We present results from a deep, coordinated \xmm+\nustar\ observation of the Seyfert
2 galaxy \iras. The \nustar\ data provide the first detection of this source in high-energy
X-rays ($E > 10$\,keV), and the broadband data show this to be a highly complex
source which exhibits relativistic reflection from the inner accretion disc, further
reprocessing by more distant material, neutral absorption, and evidence for ionised
absorption in an extreme, ultrafast outflow ($v_{\rm{out}} \sim 0.4c$). Based on
lamppost disc reflection models, we find evidence that the central supermassive black
hole is rapidly rotating ($a > 0.77$), consistent with previous estimates from the profile
of the relativistic iron line, and that the accretion disc is viewed at a fairly high
inclination ($i \sim 59$\deg). Based on extensive simulations, we find the ultrafast
outflow is detected at $\sim$4$\sigma$ significance (or greater). We also estimate that
the extreme outflow should be sufficient to power galaxy-scale feedback, and may even
dominate the energetics of the total output from the system.
\end{abstract}


\section{Introduction}

Relativistic reflection from the accretion disc is one of the primary tools at our
disposal for placing constraints on the innermost accretion geometry around black
holes. The degree of relativistic blurring can provide constraints on the inner radius
of the accretion disk, and in turn the black hole spin (\eg\ \citealt{Walton13spin};
see \citealt{Reynolds14rev} for a recent review). In addition, both the strength of the
reflected emission relative to the intrinsic continuum (the reflection fraction) and the
radial emissivity of the reflected emission from the disc can be used to constrain the
geometry and size of the primary X-ray source (the `corona'; \citealt{lightbending,
Wilkins12}). If the corona is extremely compact and very close to the black hole, the
gravitational light bending experienced by the intrinsic continuum emission can be so
strong that the reflected emission from the disc dominates the observed X-ray
spectrum. This in turn also requires a rapidly rotating black hole, such that the disk
subtends a large solid angle as seen by the X-ray source, assuming a standard thin
disc geometry (\eg\ \citealt{Parker14mrk, Dauser14}).

\iras\ is a moderately bright, nearby ($z=0.0689$) Seyfert 2 galaxy. Previous
observations with the \xmm\ (\citealt{XMM}) and \suzaku\ (\citealt{SUZAKU})
observatories revealed evidence for an extremely strong (equivalent width of
$\sim$1\,keV), relativistically broadened iron emission line (\citealt{Tan12, Ricci14}),
likely implying the presence of a rapidly rotating black hole ($a > 0.73$, where $a =
Jc/GM^{2}$ is the dimensionless spin parameter). The extreme equivalent width is
consistent with an intrinsic spectrum that is dominated by the contribution from
relativistic disc reflection, which would require an extreme accretion geometry. Based
on spectral analysis of the soft X-ray data, \cite{Ricci14} suggest that \iras\ may be an
example of an obscured analog to narrow-line Seyfert 1 galaxies (NLS1s; see
\citealt{Gallo18rev} for a recent review on their X-ray properties) in terms of its
accretion rate, \ie it may be accreting at or close-to (or even above) the Eddington limit.

Here we present results from a coordinated observation of \iras\ taken with the \nustar\
(\citealt{NUSTAR}) and \xmm\ observatories, probing for the first time the broadband
X-ray spectrum of this source. The paper is structured as follows: in Section
\ref{sec_red} we describe the observations and our data reduction procedure, and in
Section \ref{sec_spec} we present our spectral analysis. We discuss our results in
Section \ref{sec_dis} and summarise our conclusions in Section
\ref{sec_conc}.

\begin{figure*}
\begin{center}
\hspace*{-0.3cm}
\rotatebox{0}{
{\includegraphics[width=235pt]{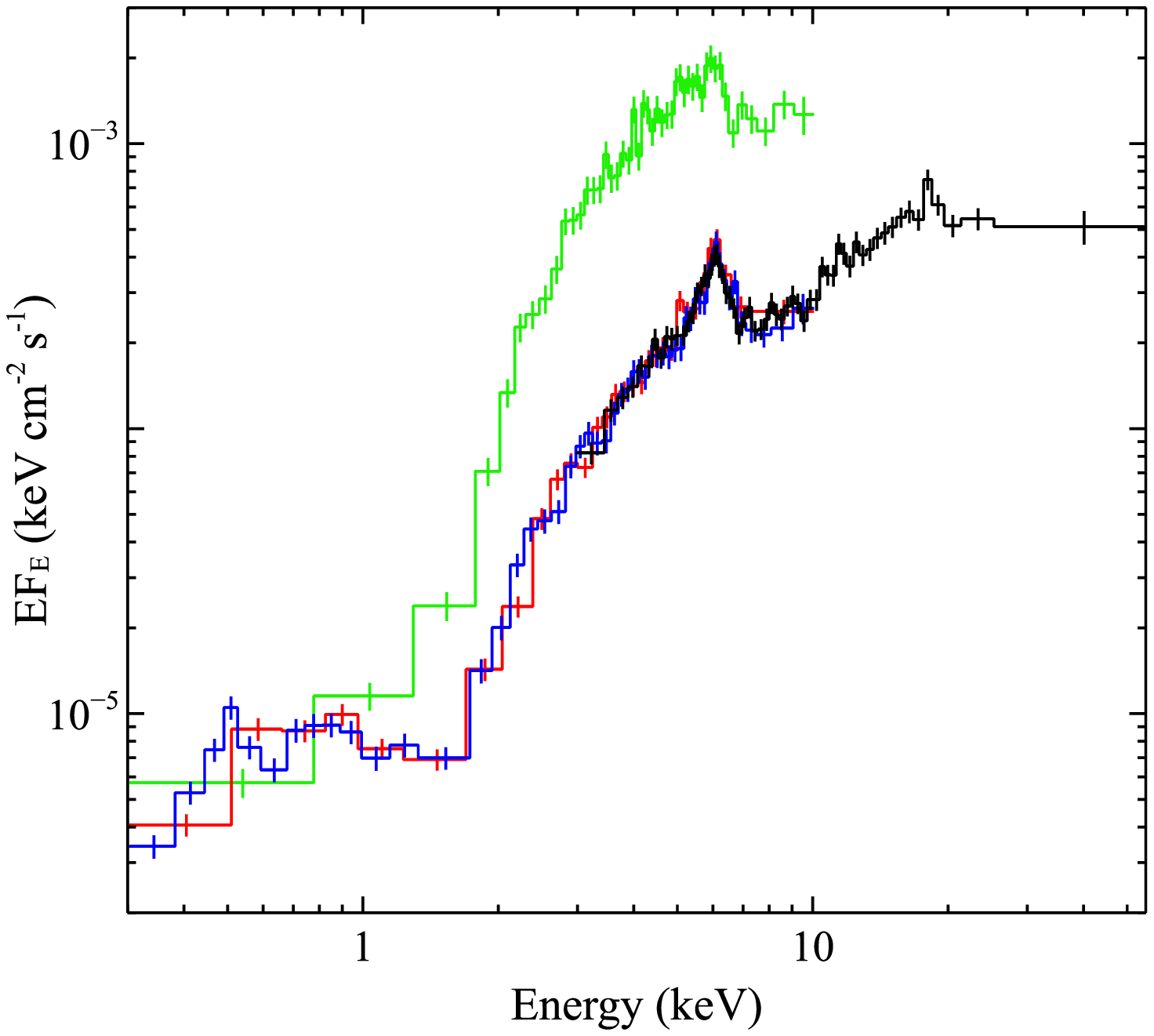}}
}
\hspace*{0.5cm}
\rotatebox{0}{
{\includegraphics[width=235pt]{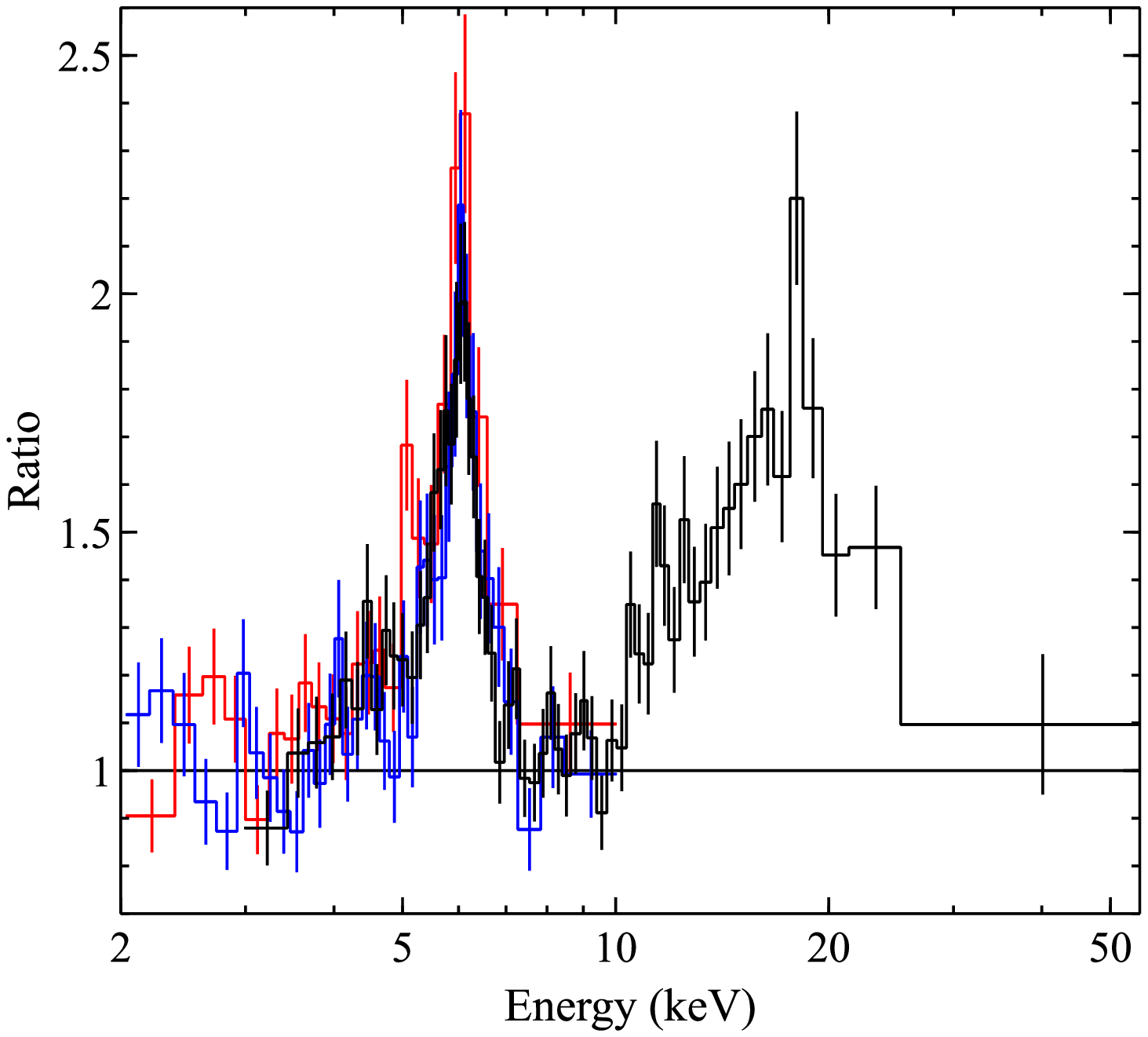}}
}
\end{center}
\vspace*{-0.3cm}
\caption{
\textit{Left panel:} the time-averaged \xmm+\nustar\ spectrum from our 2017
observation of \iras, unfolded through a model that is constant with energy. The
\nustar\ FPMA+B data are shown in black, and the \xmm\ \epicpn\ and \epicmos\ 
data are shown in blue and red, respectively. For comparison, we also show the
\epicpn\ data from 2006 (analysed in \citealt{Tan12}) in green. \textit{Right panel:}
residuals to a simple \cutoffpl\ continuum, modified by a partially covering neutral
absorber, and applied to the 2017 data over the 2--3.5, 7--10 and 40--55\,\kev\
energy ranges. The key signatures of relativistic disc reflection, \ie a broadened iron
line at $\sim$6\,\kev\ and a strong Compton hump at $\sim$20\,keV, are seen. The
data in both panels have been rebinned for visual purposes.}
\label{fig_spec}
\end{figure*}

\section{Observations and Data Reduction}
\label{sec_red}

\nustar\ and \xmm\ performed coordinated observations \iras\ in 2017; a summary of
the observations is given in Table \ref{tab_obs}. The bulk of the \nustar\ exposure was
taken contemporaneously with \xmm\ in September/October, but owing to scheduling
constraints the final segment was taken $\sim$6 weeks later. The average source flux
varied by $<$10\% in the 3--50\,keV band (see below) between the \nustar\ exposures,
and none of the individual \xmm\ or \nustar\ exposures show notable variability on
intra-observational timescales, so we treat all of the data as a single observation
despite this separation and extract a single, time-averaged broadband spectrum.

We reduced the \nustar\ data as standard using the \nustar\ Data Analysis Software
(\nustardas) v1.8.0 and instrumental calibration files from caldb v20180312. We
cleaned the unfiltered event files with \nupipeline\ using the standard depth correction,
which significantly reduces the internal high-energy background, and also excluded 
passages through the South Atlantic Anomaly (using the following settings: {\small
SAACALC}\,=\,3, {\small SAAMODE}\,=\,Optimized and {\small TENTACLE}\,=\,yes).
Source products and instrumental responses were extracted from circular regions of
radius 50$''$ using \nuproducts\ for both of the focal plane modules (FPMA/B), and
background was estimated from larger regions of blank sky on the same detector as
\iras. In order to maximise the exposure, in addition to the standard `science' (mode 1)
data, we also extracted the `spacecraft science' (mode 6) data following the procedure
outlined in \cite{Walton16cyg}. For these observations of \iras, the mode 6 data provide
$\sim$33\% of the total $\sim$400\,ks good \nustar\ exposure. As the source is
relatively faint during this epoch (see below), we combined the data from the FPMA
and FPMB modules into a single spectrum using \addascaspec, and fit the \nustar\
data over the $\sim$3--50\,keV band, above which the background dominates.

The data reduction for the \xmm\ data was also carried out following standard
procedures. We used the \xmm\ Science Analysis System (\sas v15.0.0) to clean the
raw observation files, specifically using \epchain\ and \emchain\ for the \epicpn\
detector (\citealt{XMM_PN}) and the two \epicmos\ units (\citealt{XMM_MOS}),
respectively. Source products were then extracted from the cleaned event files from
circular regions of radius 25$''$ using \xmmselect, and background was again
estimated from larger regions of blank sky on the same detector chips as \iras. These
\xmm\ observations suffered from reasonably extensive periods of background flaring,
so we utilized the method outlined in \cite{Picon04} to determine the level of
background emission that maximises the signal-to-noise (S/N) for the source for a
given energy band; as we are primarily interested in the direct emission from the AGN,
we maximise the S/N in the 5--10\,keV band. As recommended, we only extracted
single and double patterned events for \epicpn\ ({\small PATTERN}\,$\leq$\,4) and
single to quadruple patterned events for \epicmos\ ({\small PATTERN}\,$\leq$\,12),
and instrumental response files for each of the detectors were generated using
\rmfgen\ and \arfgen. We note that the observed count rates were sufficiently low that
pile-up is of no concern ($\sim$0.06\,\ctps\ and $\sim$0.02\,\ctps\ for \epicpn\ and
each \epicmos\ unit, respectively). After performing the reduction separately for the
two \epicmos\ units, we also combined these data into a single spectrum using
\addascaspec. We fit the \xmm\ data over the full 0.3--10\,keV band.

\begin{table}
  \caption{Details of the 2017 X-ray observations of IRAS\,00521-7054.}
\begin{center}
\begin{tabular}{c c c c c}
\hline
\hline
\\[-0.2cm]
Mission & OBSID & Start Date & Exposure (ks)\tmark[a] \\
\\[-0.3cm]
\hline
\hline
\\[-0.2cm]
\multirow{2}{*}{\xmm} & 0790590101 & 2017-09-30 & 65/101 \\
\\[-0.3cm]
& 0795630201 & 2017-10-02 & 55/65 \\
\\[-0.1cm]
\multirow{3}{*}{\vspace{-0.1cm} \nustar} & 60301029002 & 2017-09-30 & 106 \\
\\[-0.3cm]
& 60301029004 & 2017-10-02 & 184 \\
\\[-0.3cm]
& 60301029006 & 2017-11-17 & 111 \\
\\[-0.3cm]
\hline
\hline
\end{tabular}
\end{center}
$^{a}$ \xmm\ exposures are listed for the \epicpn/MOS detectors, after correcting for
background flaring.
\vspace*{0.3cm}
\label{tab_obs}
\end{table}

\section{Spectral Analysis}
\label{sec_spec}

We focus on a spectral analysis of the broadband \xmm+\nustar\ data, using \xspecv\
(\citealt{xspec}) to model the data. Uncertainties on the spectral parameters are quoted
at the 90\% confidence level for a single parameter of interest. Each of the datasets are
rebinned to have a minimum S/N of 5 per bin, sufficient for \chisq\ minimisation. All of
our models include a neutral absorber associated with our own Galaxy, modelled with
the \tbabs\ neutral absorption code (\citealt{tbabs}). As recommended we use the
cross-sections of \cite{Verner96} for the neutral absorption, but we adopt the
solar abundance set of \cite{Grevesse98} for self-consistency with both the \xillver\
reflection models (\citealt{xillver}) and the \xstar\ photoionisation code (\citealt{xstar}),
as these are used to model the central AGN throughout this work. The column
density of the Galactic absorption component is fixed the to $N_{\rm{H,Gal}} = 5.26
\times 10^{20}$\,\pcmsq\ (\citealt{NH}). As is standard, cross-calibration uncertainties
between the different detectors are accounted for by allowing multiplicative constants
to vary between them. We fix \epicpn\ at unity, and the others are found to be within
$\sim$10\% of unity, as expected (\citealt{NUSTARcal}).

The broadband spectrum is shown in Figure \ref{fig_spec} (left panel), along with the
previous \xmm\ data obtained in 2006 (see \citealt{Tan12}) for comparison. The data
above $\sim$2\,keV, where the direct emission from the central nucleus dominates,
shows that \iras\ was significantly fainter during our 2017 observations than the
previous X-ray observations with \xmm\ and \suzaku\ (the 2013 \suzaku\ observation
caught the source in the same flux state as the 2006 \xmm\ observations;
\citealt{Ricci14}). The observed 2--10\,keV flux in 2017 is $\sim$4 $\times
10^{-13}$\,\ergpcmsqps, a factor of $\sim$6 fainter than the previous \xmm\ and
\suzaku\ observations. Despite this, at the lowest energies (below $\sim$1\,keV) the
2006 and 2017 data show similar fluxes, implying that these energies are dominated
by diffuse plasmas and scattered emission, similar to other obscured AGN (\eg\
\citealt{Winter09, Walton18}).

To further highlight the features in the high-energy spectrum we also show the
data/model ratio of the combined \xmm+\nustar\ data above 2\,keV to a simple
absorbed \cutoffpl\ continuum, fit to the 2--3.5, 7--10 and 40--55\,keV bands
(observed frame) where the primary AGN continuum would be expected to dominate
(Figure \ref{fig_spec}, right panel). We allow the absorption to be partially covering,
but find a covering factor of $C_{\rm{f}} \sim 1$, along with a column density of
$N_{\rm{H}} \sim 7 \times 10^{22}$\,\pcmsq, a photon index of $\Gamma \sim 1.6$
and a cutoff energy of $E_{\rm{cut}} \sim 170$\,keV. A strong, broad emission
feature is clearly seen in the iron bandpass, similar to the broad iron K emission
previously reported for this source (\citealt{Tan12, Ricci14}), and a strong excess of
emission is also seen above 10\,keV. This high-energy excess peaks at $\sim$20\,keV,
as expected for a Compton reflection continuum. In addition to these features, a
narrower core to the iron emission at 6.4\,keV is clearly visible.

Interestingly, the column density we find when using this simple model is similar to
the neutral columns inferred with similar models in both \cite{Tan12} and
\cite{Ricci14} ($\sim$6 and $\sim$7 $\times 10^{22}$\,\pcmsq, respectively),
suggesting that the low flux observed here is not related to strong changes in the
line-of-sight absorption. Instead, the flux variability is likely intrinsic to the
source, and related to changes in the accretion rate through the inner regions of the
disc.

\subsection{Broadband Continuum Modeling (Model 1)}
\label{sec_cont}

We construct a model for the broadband continuum consisting of the primary
Comptonised X-ray continuum, relativistic reflection from the inner accretion disc to
account for the broad iron emission, a partially covering neutral absorber associated
with the nucleus and a fully covering neutral absorber to account for the galaxy-scale
column in the \iras\ galaxy (similar to the Galactic column), a more distant reflector to
account for the narrower iron emission, and a collisionally ionised plasma to
additionally account for the constant soft X-ray emission. Both of the neutral
absorbers associated with \iras\ are again modeled with the \tbabs\ absorption code,
and are assumed to be at the redshift of the host-galaxy. Relaxing this assumption
for the nuclear absorber (refered to as \tbabs$_{2}$) does not improve the fit, and the
X-ray constraints on the absorber redshift are consistent with that of the host galaxy.
This absorption component is allowed to be partially covering to account for the weak
scattered nuclear emission ubiquitously seen in the soft X-ray band (in addition to
ionized plasma emission) in absorbed AGN (\eg\ \citealt{Winter09}). The nuclear
absorber only acts on the direct emission from the central nucleus (the primary X-ray
continuum and the relativistic disc reflection), while both of the absorbers associated
with the galaxy-scale absorption in \iras\ (\tbabs$_{1}$) and our own Galaxy
(\tbabs$_{\rm{Gal}}$) act on all model components.

For the relativistic reflection, we use the \relxilllpcp\ model (v1.2.0; \citealt{relxill}). This
accounts for both the continuum emission from the illuminating X-ray source (assuming
an \nthcomp\ continuum, parameterised by a photon index, $\Gamma$, and the
electron temperature, \kte; \citealt{nthcomp1, nthcomp2}) and the reflected emission
from the accretion disc. The disc reflection contribution is computed self-consistently
assuming a simple lamppost geometry with a thin disc, both in terms of the emissivity
profile of the disk and the strength of the reflected emission (the reflection fraction,
\Rfrac; see \citealt{relxill_norm}). We assume that the inner accretion disc
reaches the innermost stable circular orbit (ISCO) in all our analysis, and fix the outer
disk to the maximum value allowed by the model (1000\,\rg), so both the emissivity
profile and the reflection fraction are set by the dimensionless spin of the black hole,
$a$, and the height of the illuminating X-ray source, $h$. The height of the corona is fit
in units of the vertical horizon radius to ensure that the X-ray source is always outside
this point, but where relevant we convert this to units of \rg\ when quoting the results.
The other key free parameters are the inclination, the ionisation parameter, and the iron
abundance of the disk ($i$, $\xi$ and $A_{\rm{Fe}}$, respectively; the rest of the
cosmically abundant elements are assumed to have solar abundances). The
ionisation parameter is defined as $\xi = L_{\rm{ion}}/nR^{2}$, where $L_{\rm{ion}}$ is
the ionising luminosity (assessed between 1--1000\,Ry), $n$ is the density of the
material, and $R$ is the distance between the material and the ionising source.

\begin{figure*}
\begin{center}
\hspace*{-0.25cm}
\rotatebox{0}{
{\includegraphics[width=395pt]{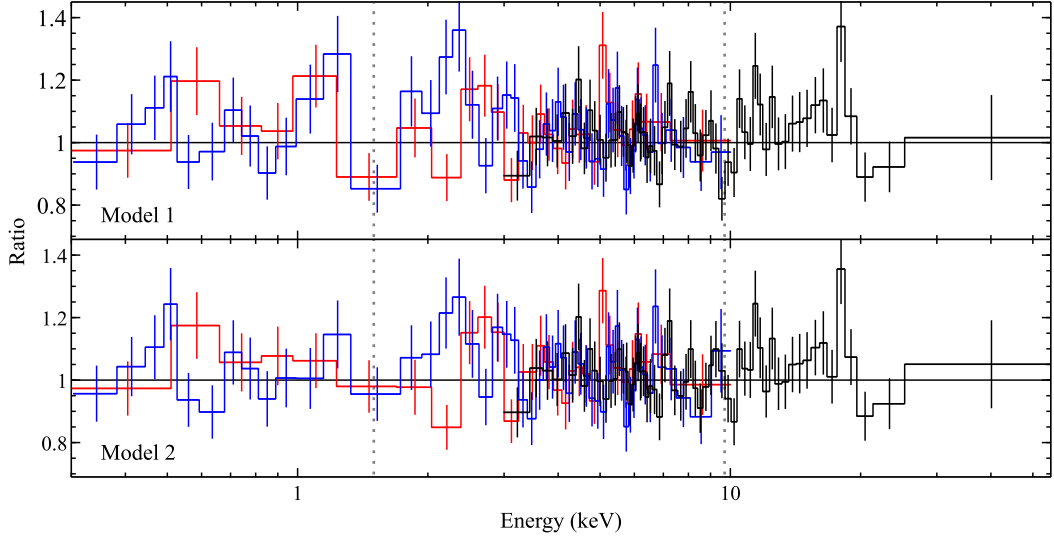}}
}
\end{center}
\vspace*{-0.3cm}
\caption{
Data/model ratios for our baseline broadband continuum model (Model 1, see Section
\ref{sec_cont}), and the model including strongly blueshifted ionised absorption (Model
2, see Section \ref{sec_ufo}). The \nustar\ and \xmm\ data are shown in the same
colours as in Figure \ref{fig_spec}, and the data have again been rebinned for visual
clarity. The vertical dotted lines show the approximate energies of the main features
of the ionised, blueshifted absorber.
}
\label{fig_ratio}
\end{figure*}

For the more distant reflection we use the \xillvercp\ model (configured such that this
component only provides a reflection spectrum), as this also assumes an \nthcomp\
ionising continuum. Although this implicitly assumes a slab geometry for the distant
reflector, and the true geometry may more closely resemble an equatorial torus, our
recent work on the absorbed AGN in IRAS\,13197-1627 found that similar results
were obtained regardless of the geometry assumed for the distant reflector
(\citealt{Walton18}).\footnote{Although direct comparisons with the available
torus models (\eg\ \borus; \citealt{borus}) are not straightforward, owing to the
different parameterizations of the input continuum assumed (for example, at the time
of writing \borus\ assumes a powerlaw with an exponential high-energy cutoff, which
has known differences to the more physical Comptonized continuum adopted here;
see \eg\ \citealt{Zdziarski03}, \citealt{Fabian15}), simple tests using \borus\ instead
of \xillvercp\ give similar results for the key disc reflection parameters to those
presented here.} \xillvercp\ shares most of its key free parameters with the
\relxilllpcp\ model (aside from those associated with the relativistic blurring), and we
assume that the distant reflector is neutral. While the core of the iron emission is
clearly narrower than the relativistic emission, modeling it with a simple Gaussian
profile suggests this emission does still have some width ($\sigma \sim 0.18$\,keV).
This is consistent with the results presented by \cite{Gandhi15}, who find that the
majority of the narrow core of the iron emission in Seyfert galaxies arises from scales
interior to the dust sublimation radius, and is likely associated with the broad line
region (see also \citealt{Miller18}). We therefore assume the line width here arises
due to velocity broadening, so we convolve the \xillver\ component with a Gaussian
kernel (\gsmooth) which has a constant ratio of $\sigma/E$; the width of this Gaussian
kernel is evaluated at 6\,keV. Finally, we account for the distant plasma emission with
a \mekal\ model. The formal model expression is as follows: \tbabs$_{\rm{Gal}} 
\times$\tbabs$_{1} \times$ (\gsmooth$\otimes$\xillvercp\ + \mekal\ + \tbabs$_{2}
\times$\relxilllpcp).

\begin{table*}
  \caption{Results obtained for the free parameters in the lamppost
  reflection models fit to the broadband \xmm+\nustar\ data for \iras.}
\begin{center}
\begin{tabular}{c c c c c}
\hline
\hline
\\[-0.2cm]
Model Component & \multicolumn{2}{c}{Parameter} & \multicolumn{2}{c}{Model} \\
\\[-0.35cm]
& & & 1 & 2 \\
\\[-0.3cm]
\hline
\hline
\\[-0.1cm]
\tbabs$_{1}$ (galaxy-scale) & $N_{\rm{H,1}}$ & [$10^{20}$ cm$^{-2}$] & $<6.2$ & $4.5^{+3.5}_{-4.1}$ \\
\\
\tbabs$_{2}$ (nuclear) & $N_{\rm{H,2}}$ & [$10^{22}$ cm$^{-2}$] & $6.8^{+2.2}_{-0.4}$ & $6.7^{+2.2}_{-0.6}$ \\
\\[-0.3cm]
& $C_{\rm{f,2}}$ & [\%] & $96.4^{+1.6}_{-0.7}$ & $96.3^{+0.7}_{-1.7}$ \\
\\
\relxilllpcp\ & $\Gamma$ & & $1.96^{+0.06}_{-0.09}$ & $1.91^{+0.10}_{-0.17}$ \\ 
\\[-0.3cm]
& $kT_{\rm{e}}$\tmark[a] & [keV] & $36^{+12}_{-21}$ & $43^{+24}_{-32}$ \\
\\[-0.3cm]
& $a$ & & $>0.73$ & $>0.77$ \\
\\[-0.3cm]
& $i$ & [\deg] & $63^{+3}_{-27}$ & $59^{+3}_{-16}$ \\
\\[-0.3cm]
& $h$ & [\rg] & $<7$ & $<5$ \\
\\[-0.3cm]
& \Rfrac\tmark[b] & & $>1.6$ & $>1.7$ \\
\\[-0.3cm]
& $\log\xi$ & $\log$[\ergcmps] & $1.1^{+1.9}_{-0.8}$ & $2.1^{+0.4}_{-2.0}$ \\
\\[-0.3cm]
& $A_{\rm{Fe}}$\tmark[c] & [solar] & $3.1^{+0.4}_{-1.6}$ & $3.2^{+0.6}_{-0.9}$ \\
\\[-0.3cm]
& $\log[\rm{Norm}]$ & & $-3.69^{+0.34}_{-1.88}$ & $-3.32^{+0.22}_{-1.98}$ \\
\\
\xstar\ & $\log\xi$ & $\log$[\ergcmps] & -- & $4.7 \pm 0.1$ \\
\\[-0.3cm]
& \nh\ & [$10^{22}$ cm$^{-2}$] & -- & $2.8^{+1.7}_{-1.1}$ \\
\\[-0.3cm]
& $z_{\rm{abs}}$ & & -- & $-0.349 \pm 0.009$ \\
\\
\xillvercp\ & $\sigma$ (at 6\,keV) & [keV] & $0.19^{+0.06}_{-0.05}$ & $0.20 \pm 0.06$ \\
\\[-0.3cm]
& Norm & [$10^{-6}$] & $4.3^{+1.4}_{-0.7}$ & $4.4^{+1.1}_{-0.8}$ \\
\\
\mekal\ & $kT$ & [keV] & $0.76^{+0.06}_{-0.11}$ & $0.79^{+0.09}_{-0.10}$ \\
\\[-0.3cm]
& Norm & [$10^{-6}$] & $1.5^{+1.4}_{-0.4}$ & $1.1^{+0.7}_{-0.4}$ \\
\\[-0.2cm]
\hline
\\[-0.25cm]
\chisq/DoF & & & 621/553 & 585/550 \\
\\[-0.25cm]
\hline
\hline
\end{tabular}
\label{tab_param}
\end{center}
\flushleft
$^a$ $kT_{\rm{e}}$ is quoted in the rest-frame of the X-ray source (\ie prior to any
gravitational redshift), based on the best-fit lamppost geometry. \\
$^b$ \Rfrac\ is calculated self-consistently for the lamppost geometry from
$a$ and $h$; the errors represent the range of values permitted by varying these
parameters within their 90\% uncertainties. The maximum value permitted by the
self-consistent RELXILLLPCP model is \Rfrac\ $\sim$ 20.\\
$^c$ The iron abundance is linked across all spectral components associated
with the nucleus of \iras. \\
\vspace{0.4cm}
\end{table*}

For self-consistency within our model, we make sure to link the iron abundance
parameters across all the different model components associated with \iras, and we
also assume their abundances for the other cosmically abundant elements are solar.
We also link the photon index between the \relxilllpcp\ and \xillvercp\ components.
However, in the later versions of \relxill\ (v1.0.4 onwards) the electron temperature is
given in the rest-frame of the X-ray source for the lamppost models, prior to any
gravitational redshift ($z_{\rm{grav}}$) that should be applied to the emission as seen
by a distant observer. As such, we apply this redshift to the rest-frame electron
temperature when determining the illuminating spectrum seen by the distant reflector.
This depends on both the spin of the black hole ($a$) and the height of the X-ray
source ($h$; here in units of \rg) following equation \ref{eqn_zgrav}.

\vspace*{-0.1cm}
\begin{equation}
(1+z_{\rm{grav}}) = \bigg(1-\frac{2h}{h^{2}+a^{2}}\bigg)^{-\frac{1}{2}}
\label{eqn_zgrav}
\end{equation}

This model (which we refer to as Model 1) describes the broadband spectral shape
of \iras\ well, with \chisq\ = 621 for 553 degrees of freedom (DoF). The best-fit
parameters are given in Table \ref{tab_param}, and we show the data/model ratio in
Figure \ref{fig_ratio} (top panel). Even when allowing for partially covering absorption,
the data prefer a large reflection fraction of \Rfrac\ $>$ 1.6. This requires strong
gravitational lightbending for a standard thin accretion disc (which would otherwise
give \Rfrac\ $\sim$ 1). In turn, this requires both a rapidly rotating black hole and a
relatively compact illuminating corona (\citealt{lightbending, relxill_norm}), and we find
$a > 0.73$ and $h < 7$\,\rg. Our results also imply that \iras\ has a super-solar iron
abundance ($A_{\rm{Fe}} \sim 3$), and that we view the accretion disk at a
moderately high inclination ($i \sim 60$\deg). We stress that, with the exception of the 
galaxy-scale absorber associated with \iras, the removal of any of the broadband
continuum components included in this model significantly degrades the fit, increasing
the \chisq\ by $\geq$8 per free parameter. The galaxy-scale absorber is not formally
required by the data (hence only an upper limit on the column is obtained). However,
we retain this component in our final model to account for more realistic errors on the
photon index, which is influenced in part by the slope of the spectrum at the lowest
energies ($\lesssim$1\,keV) where the scattered continuum contributes.

\subsection{Ionised Absorption (Model 2)}
\label{sec_ufo}

Although Model 1 describes the broadband continuum well, the \nustar\ data show
evidence for an absorption feature at $\sim$9.5\,keV in the observed frame\footnote{At
this energy the \xmm\ S/N is sufficiently low that these data are not sensitive to atomic
line features, but are consistent with \nustar.}. There are no known instrumental
features close to this energy. Adding a Gaussian absorption feature provides a
reasonable improvement of $\Delta\chi^{2} = 14$ for three additional free parameters,
and we find a rest-frame line energy of $E = 10.1^{+0.2}_{-0.1}$\,keV, a line width of
$\sigma < 0.6$\,keV (corresponding to a velocity broadening of $\lesssim
18,000$\,\kmps), and an equivalent width of $120^{+180}_{-50}$\,eV. Given
the systemic redhift of \iras, the line energy implies a blueshift of either $z_{\rm{abs}}
\sim 0.31$ or 0.34 relative to the cosmological redshift of \iras, assuming an association
with \fexxvi\ or \fexxv, respectively, either of which would place the absorber firmly in the
`ultrafast' outflow (UFO) category (\eg\ \citealt{Tombesi10a, Tombesi10b, Gofford13,
Nardini15, Matzeu17, Parker18ufo}), and would even make it one of the most extreme
in terms of the observed blueshift.

\begin{figure*}
\begin{center}
\hspace*{-0.25cm}
\rotatebox{0}{
{\includegraphics[width=395pt]{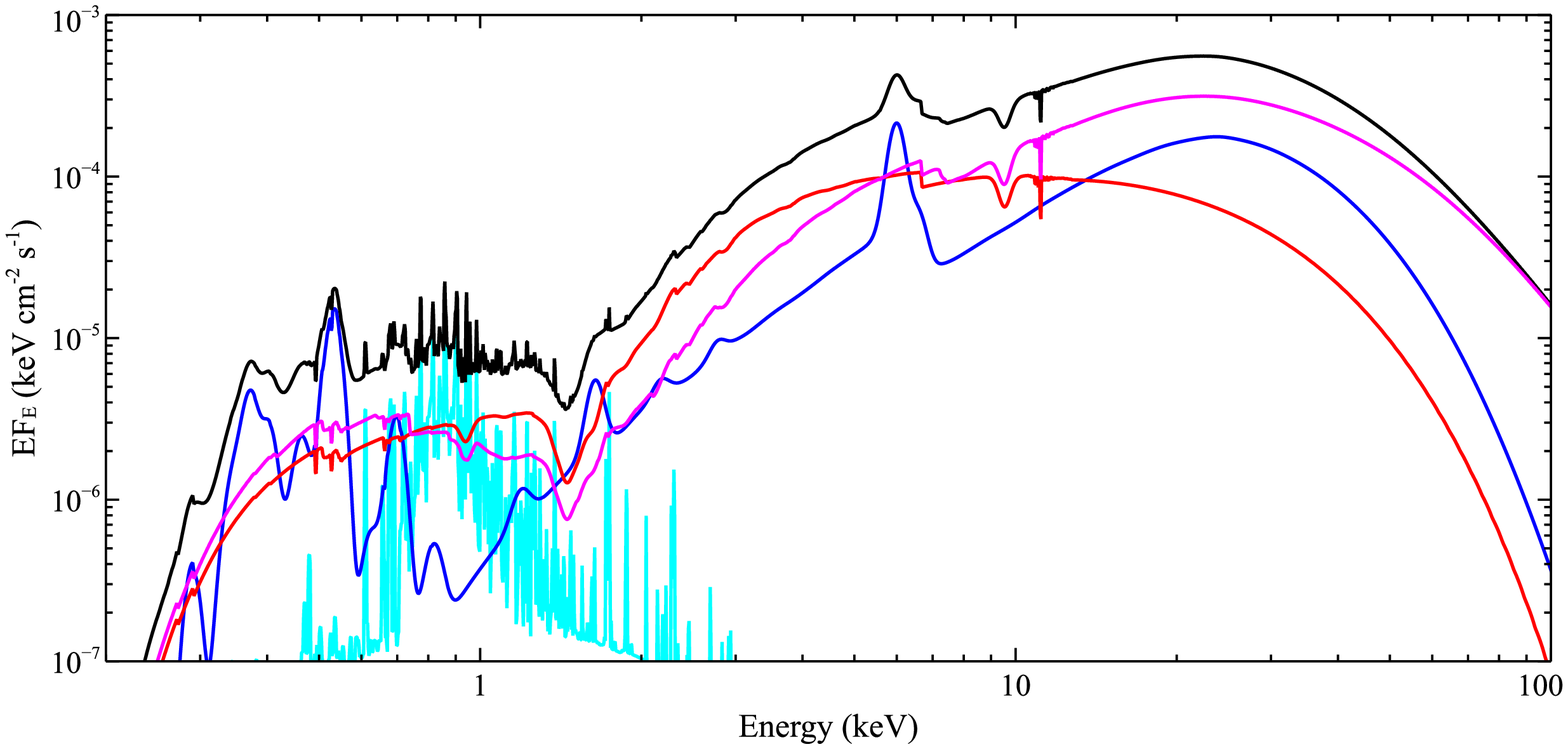}}
}
\end{center}
\vspace*{-0.3cm}
\caption{
The relative contributions of the different components for our final model including the
highly blueshifted ionized absorption (Model 2). The total model is shown in black,
the Comptonised continuum in red, the relativistic disc reflection in magenta, the distant
reflection in blue, and the distant plasma in cyan.
}
\label{fig_mod}
\end{figure*}

\begin{figure*}
\begin{center}
\hspace*{-0.25cm}
\rotatebox{0}{
{\includegraphics[width=235pt]{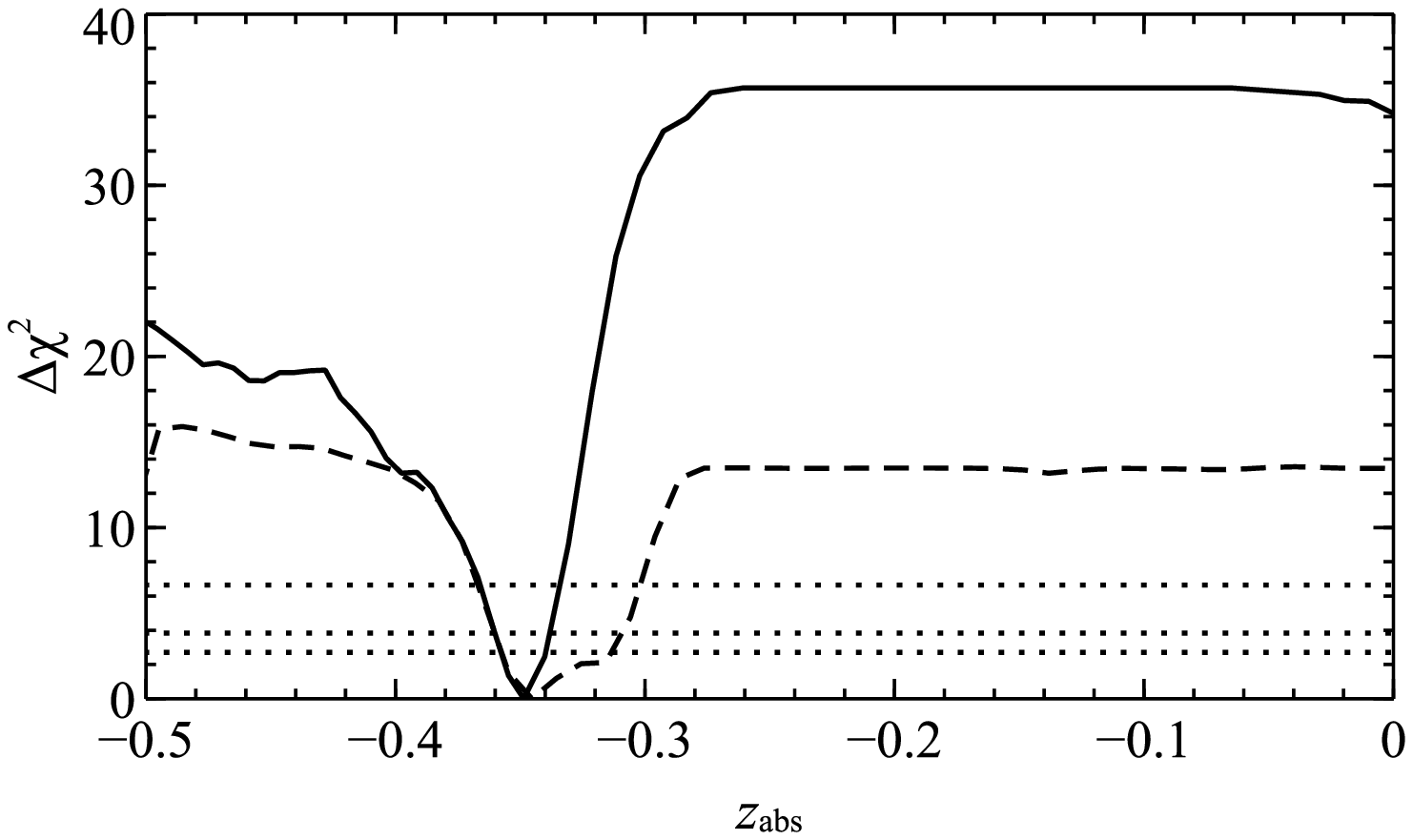}}
}
\hspace{0.75cm}
\rotatebox{0}{
{\includegraphics[width=235pt]{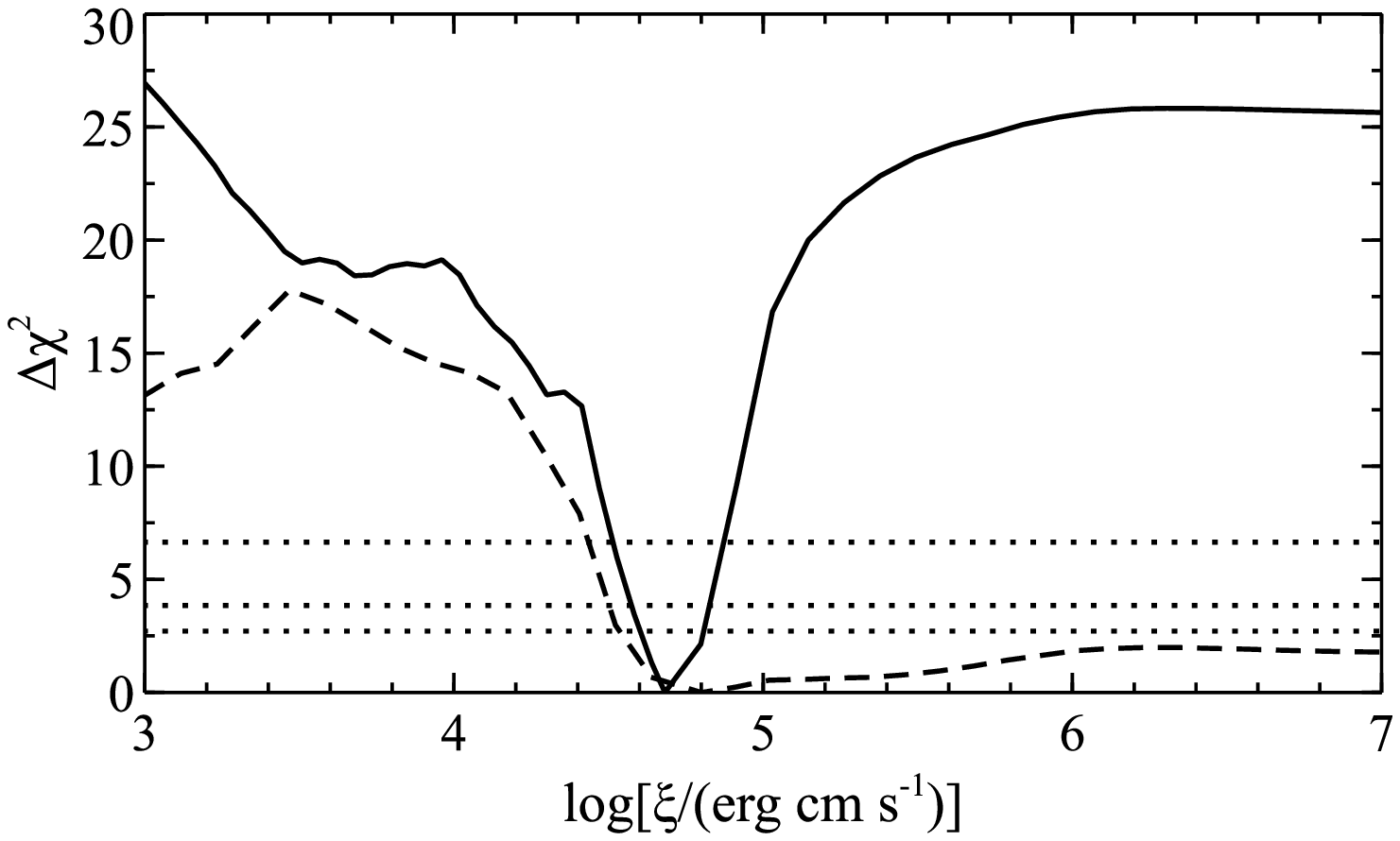}}
}
\end{center}
\caption{
The $\Delta$\chisq\ confidence contours for the blueshift (left panel, given in the rest
frame of \iras) and the ionisation parameter (right panel) of the highly ionised absorber.
The horizontal dotted lines represent the 90, 95 and 99\% confidence levels for a
single parameter of interest. The solid curves show the confidence contours for the full
dataset, and the dashed curves show the contours after excluding the 1.25--1.65\,keV
energy range to cut out the feature at $\sim$1.5\,keV.
}
\label{fig_zabs}
\end{figure*}

To investigate this further we replace the Gaussian absorption line with a physical
photoionisation model using \xstar. We use a grid of absorption models with the
ionisation parameter, column density and iron abundance as free parameters. All
other elements have solar abundances, and these absorption models also assume a
velocity broadening of 10,000\,\kmps\ (through the `turbulent' velocity parameter in
\xstar; $v_{\rm{turb}}$), based on the constraints above and the line broadening seen
in other well-studied UFO sources (\eg\ \citealt{Pounds03, Nardini15}). During our
analysis, the iron abundance is linked to that of the continuum model components,
and as with the neutral absorbers, this absorption component is only applied to the
direct emission from the central nucleus (the \relxilllpcp\ component), such
that the model expression is updated to the following: \tbabs$_{\rm{Gal}}
\times$\tbabs$_{1} \times$(\gsmooth$\otimes$\xillvercp\ + \mekal\ + \tbabs$_{2}
\times$\xstar$\times$ \relxilllpcp).

Since the ionisation parameter is defined using the ionising luminosity across the
1--1000\,Ry bandpass, the intrinsic broadband (optical to X-ray) spectral energy
distribution (SED) actually plays an important role in setting the ionisation parameter
at which highly ionised iron transitions are observed. We therefore inspected the data
from the Optical Monitor on board \xmm\ (\citealt{XMM_OM}), which took exposures
in each of its optical--UV filters (V, B, U, UVM2, UVW1 and UVW2). However, while
\iras\ is clearly detected in each of these filters, in almost all cases the OM data show
an extended counterpart, implying a non-negligible contribution from the host galaxy.
Given this, and the fairly heavy obscuration towards the nucleus, we conclude that it
is beyond the scope of this work to observationally determine the intrinsic SED for the
central AGN for use with \xstar. Instead, we take a simpler approach, and for the input
to \xstar\ we assume an SED in which the X-ray emission is dominated by
Comptonisation, and the optical/UV emission is dominated by a standard accretion
disc (\citealt{Shakura73}). The X-ray continuum is modeled with \nthcomp, with the
spectral parameters ($\Gamma$, $kT_{\rm{e}}$) set by the continuum analysis above, 
and the accretion disk is modeled with \diskbb\ (\citealt{diskbb}). To set the disc
temperature, based on the arguments in \cite{Ricci14} we assume that \iras\ is 
accreting at roughly its Eddington rate during the higher flux observations in the
archive. Adopting a bolometric correction for the 2--10\,keV band of $\kappa_{2-10}
\equiv L_{2-10}/L_{\rm{bol}} = 150$ -- as appropriate for high-Eddington sources
(\citealt{Vasudevan09, Lusso10}) -- implies an intrinsic bolometric luminosity of
$L_{\rm{bol}} \sim 6 \times 10^{45}$\,\ergps\ during these epochs (\citealt{Tan12}),
given the luminosity distance of 300\,Mpc (based on a standard cosmology with
$H_{0}$ = 73\,\H0, $\Omega_{\rm{matter}} = 0.27$ and $\Omega_{\rm{vacuum}} =
0.73$). In turn, this implies a black hole mass of $M_{\rm{BH}} \sim 5 \times 10^{7}$
\msun, and therefore an inner disc temperature of  $T_{\rm{in}} \sim 0.05$\,keV.
Finally, we set the normalisation of the disc relative to the X-ray continuum to give the
same bolometric correction as above.

\begin{figure}
\begin{center}
\hspace*{-0.25cm}
\rotatebox{0}{
{\includegraphics[width=235pt]{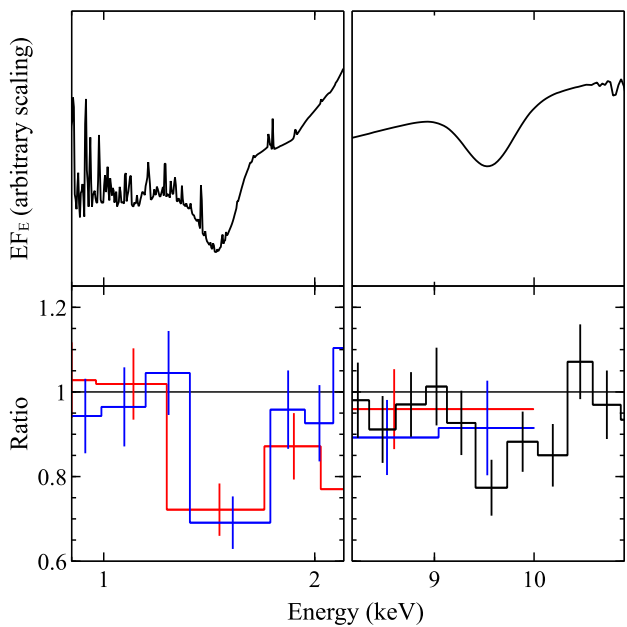}}
}
\end{center}
\caption{
Zoom-in around the key absorption features for the best-fit \xstar\ model (Model 2,
shown in full in Figure \ref{fig_mod}; top panels) and the data/model ratios after the
\xstar\ component has been removed (without refitting; bottom panels). The left panels
focus on the feature at $\sim$1.5\,keV, and the right panels focus on the feature at
$\sim$9.5 keV. For the model plots, we only show the total model (rather than the
individual components) for clarity, and the colours in the ratio plots have the same
meaning as in Figures \ref{fig_spec} and \ref{fig_ratio}.
}
\label{fig_lines}
\end{figure}

The inclusion of the photoionised absorber provides a much larger improvement to
our baseline continuum model than the single Gaussian absorption line, with
$\Delta\chi^{2} = 36$ for three additional free parameters. We refer to this as Model
2 and, given the level of improvement, consider this our preferred model (see Section
\ref{sec_sims} for a formal assessment of the significance of this outflow); the best-fit
parameter values are given in Table \ref{tab_param}, the data/model ratio for the fit is
shown in Figure \ref{fig_ratio} (bottom panel), and we show the relative contributions of
the different components for this model in Figure \ref{fig_mod}. Figure \ref{fig_zabs}
shows the confidence contours for both the blueshfit (relative to the cosmological
redshift of \iras) and the ionisation parameter of the ionised absorber (solid lines); the
data clearly prefer the \fexxv\ solution (which corresponds to $z_{\rm{abs}} \sim 0.35$)
over the \fexxvi\ solution (which would have $z_{\rm{abs}} \sim 0.31$). This is because,
with the \fexxv\ solution, the photoionised absorber is also able to match a second
absorption feature at $\sim$1.5\,keV (observed frame) in addition to the high-energy
feature at $\sim$9.5\,keV (see Figure \ref{fig_ratio}, and also Figure \ref{fig_lines}
where we show a zoom-in on these two features). This $\sim$1.5\,keV feature is also
produced by iron in the model, in this case the complex of L-shell transitions from Fe
{\small XXI-XXIV}, which matches the observed line energy for the same blueshift as
the \fexxv\ solution for the line observed at 9.5\,keV, resulting in a much stronger total
statistical improvement than the single Gaussian. Accounting for the relativistic
corrections necessary for such extreme blueshifts, we find the line-of-sight velocity of
the absorber to be $\beta_{\rm{LoS}} = v_{\rm{LoS}}/c = -0.405 \pm 0.012$.

Given the presence of a photoionised absorber, we also test for the presence
of associated photoionised emission. We again use \xstar, assuming the same
input continuum, ionisation state, column density and iron abundance as the absorber.
The emitter is placed at the redshift of \iras, and we crudely attempt to account for the
expected broadening for a diverging outflow based on the observed outflow velocity
using another Gaussian smoothing kernel. However, we find that the data are not
sensitive to this emission; adding such a component provides a negligible improvement
in the fit ($\Delta\chi^{2} < 1$), and so we do not include this in our final model. In
principle the normalisation of the \xstar\ emission component, $\kappa$, can be used
to determine the solid angle subtended by the wind, as $\Omega = \kappa
D_{\rm{kpc}}/L_{38}$, where $\Omega$ is the solid angle (normalised by $4\pi$ such
that $0 \leq \Omega \leq 1$), $D_{\rm{kpc}}$ is the distance in kpc, and $L_{38}$ is
the ionising luminosity in units of $10^{38}$\,\ergps\ (csee \eg\ \citealt{Reeves18b}).
However, taking the values for $D_{\rm{kpc}}$ and $L_{38}$ discussed above, in this
case the limits the current data can place on the normalisation are sufficiently weak
that $\Omega$ is completely unconstrained.

\begin{figure}
\begin{center}
\hspace*{-0.25cm}
\rotatebox{0}{
{\includegraphics[width=235pt]{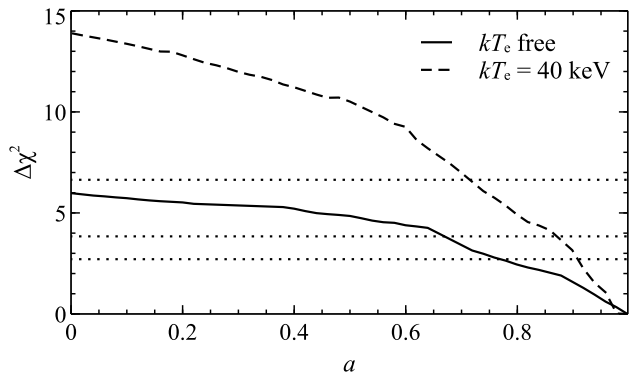}}
}
\end{center}
\vspace*{-0.3cm}
\caption{
The $\Delta$\chisq\ confidence contours for the spin parameter from our preferred
model for the broadband data of \iras\ (Model 2, see Section \ref{sec_ufo}). We show
contours computed with the electron temperature free to vary (solid line) and
assuming a temperature of $kT_{\rm{e}} = 40$\,keV (dashed line). The horizontal
dotted lines represent the 90, 95 and 99\% confidence levels for a single parameter of
interest.}
\label{fig_spin}
\end{figure}

\begin{figure}
\begin{center}
\hspace*{-0.25cm}
\rotatebox{0}{
{\includegraphics[width=235pt]{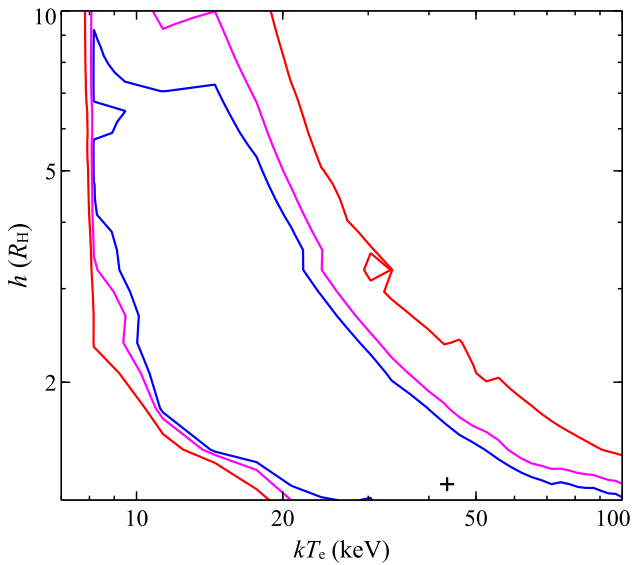}}
}
\end{center}
\vspace*{-0.3cm}
\caption{
2D confidence contours for $h$ and $kT_{\rm{e}}$ for \iras\ (using Model 2). The 90,
95 and 99\% confidence contours for 2 parameters of interest are shown in blue,
magenta and red, respectively. Note that here, $h$ is given in units of the vertical
horizon radius, \rh. A strong degeneracy is seen between these two parameters, as
$kT_{\rm{e}}$ is given in the rest-frame of the X-ray source and $h$ plays an
important role in setting the gravitational redshift (Equation \ref{eqn_zgrav}). This
degeneracy exists down to the temperature at which, in the limit of no gravitational
redshift, the corona would be unable to reproduce the observed hard X-ray flux
(which occurs at $kT_{\rm{e}} \sim 8$\,keV).}
\label{fig_kTe}
\end{figure}

The key continuum parameters are all consistent with the values obtained in Model 1,
and the constraints are generally similar. Again we find the model prefers a large
reflection fraction, and thus a rapidly rotating black hole ($a > 0.77$). We show the
confidence contour for the black hole spin in Figure \ref{fig_spin}; although a high spin
is preferred, the level at which a non-rotating black hole is excluded is not particularly
strong. However, it is worth noting that the best-fit solutions at low spin also require
a very low temperature for the corona of $kT_{\rm{e}} \sim 10$\,keV. This is because
of the complex interplay that exists between some of the main continuum parameters
in our model; the electron temperature is degenerate with both the black hole spin and,
in particular, the height of the corona. This is because the temperature is assessed in
the rest-frame of the X-ray source (rather than the observed frame) and these
parameters set the gravitational redshift (Equation \ref{eqn_zgrav}). We show a 2-D
confidence contour showing the degeneracy between $h$ and $kT_{\rm{e}}$ in Figure
\ref{fig_kTe}. This degeneracy exists down to the point at which, in the limit of negligible 
gravitational redshift, the temperature becomes too low to produce the observed hard
X-ray flux, at which point the fit quickly degrades. While coronae with low temperatures
have been reported in a few rare cases (\citealt{Tortosa17, Kara17}), intrinsic
temperatures are typically $kT_{\rm{e}} \sim 40-50$\,keV (even after correcting
approximately for gravitational redshift; \citealt{Fabian15}), similar to the best-fit value
found here. We therefore also re-compute the confidence contour for the spin with the
electron temperature fixed to $kT_{\rm{e}} = 40$\,keV, which we also show in Figure
\ref{fig_spin} (dashed line). Unsurprisingly, the constraint on the spin is much tighter
($a > 0.91$; repeating this analysis with Model 1 also gives similar results). This is
because the model is now unable to lower the electron temperature when in regions of
parameter space in which the gravitational redshift would be weaker (lower $a$, higher
$h$) to reproduce the strong curvature observed above $\sim$10\,keV, and thus the
preference for a large reflection fraction -- and in turn a high spin -- is even stronger.

\subsubsection{Statistical Simulations}
\label{sec_sims}

To formally test the statistical significance of the ionised outflow, we perform a series
of spectral simulations. Using the same response and background files, and adopting
the same exposure times and relative extraction areas as the real data used here, we
simulated \nsims\ sets of \xmm\ (pn, combined MOS1 and MOS2) and \nustar\
(combined FPMA and FPMB) spectra with the {\small FAKEIT} command in XSPEC
based on the best-fit model without the \xstar\ grid (Model 1), allowing for independent
Poisson fluctuations on both the simulated source and background spectra. Each of
the simulated datasets was background subtracted and rebinned in the same manner
as the real data, and analysed over the same bandpass. We initially fit each of the
combined datasets with Model 1, before adding the \xstar\ grid in order to determine
the improvement in \chisq\ this extra model component provides by chance, linking the
iron abundance to that of the other model components (as in our analysis of the real
data). To account for the number of trials we scan the absorber velocity between
$\beta_{\rm{LoS}} = 0$ and 0.5 in 50 steps, and then run a full error scan on the key
wind parameters for the best fit found to determine the maximum $\Delta\chi^{2}$
improvement provided. Of the \nsims\ datasets simulated, \ngtr\ returned a chance
improvement equivalent to or greater than that observed (at any velocity searched);
we show the $\Delta\chi^{2}$ distribution in Figure \ref{fig_sims}. This implies that the
outflow seen in the real data is significant at close to (or above) the 4$\sigma$ level
(the low-number statistics associated with having no simulations that give a false
positive mean the real probability could still be even larger). We stress that the
simulations undertaken in this work also allow for a suitably complex continuum
including reflection, as necessary to robustly determine the significance of the
absorption (see the discussion in \citealt{Zoghbi15}).

\subsubsection{The Nature of the 1.5\,keV Feature}

Some caution may still be required, as the background for the \xmm\ EPIC detectors
is known to contain fluorescent emission from aluminium close in energy to the
absorption feature inferred at $\sim$1.5\,keV. In addition, this feature is also close in
energy to the band in which the \mekal\ plasma component, which includes line
emission from a variety of different atomic species resulting in a complex low-energy
spectrum, makes the strongest contribution. We therefore performed a variety of tests
to determine whether the statistical detection of the ionised absorber could be driven
by any of these issues, and investigate the potential nature of the $\sim$1.5\,keV
feature in more detail.

\begin{figure}
\begin{center}
\hspace*{-0.3cm}
\rotatebox{0}{
{\includegraphics[width=235pt]{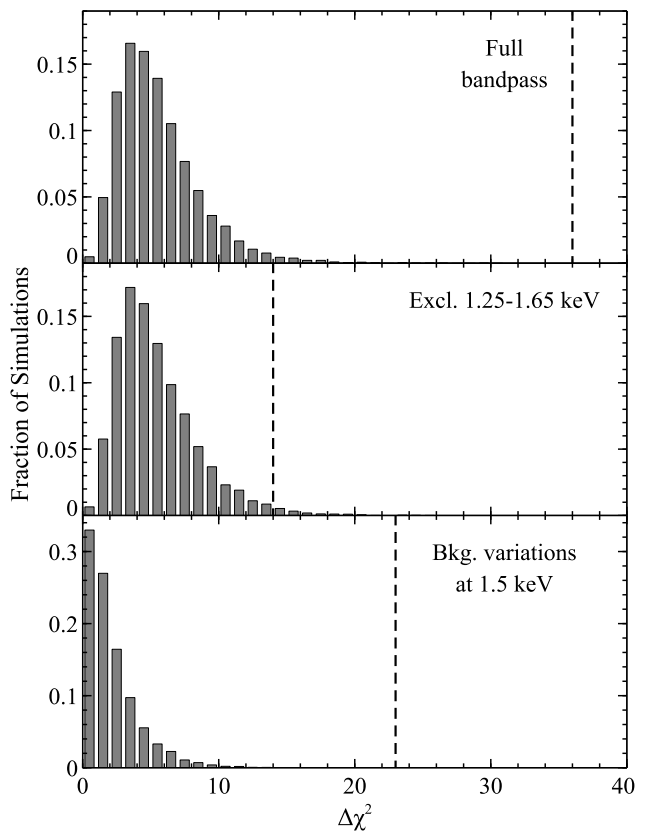}}
}
\end{center}
\vspace*{-0.3cm}
\caption{
The $\Delta\chi^{2}$ distributions from our various analyses of the 10,000 simulations
performed to assess the significance of the ultrafast outflow in \iras. The top panel
shows the analysis with the full bandpass for each simulation, the middle panel shows
the analysis excluding the 1.25--1.65\,keV energy range (testing the significance of
just the Fe K absorption), and the bottom panel shows the analysis of the background
variations at 1.5\,keV (testing the nature of the Fe L absorption). In each case, the
vertical dashed line shows the $\Delta\chi^{2}$ obtained with the real data for
comparison.}
\label{fig_sims}
\end{figure}

First, to be conservative we also repeat our analysis on both the real and the
simulated data after excluding the 1.25--1.65\,keV energy range from the \xmm\
detectors. We find that the addition of the \xstar\ component improves the fit in the
real data by $\Delta\chi^{2} = 14$ for three additional free parameters (similar to
the improvement provided by the single Gaussian line). The confidence contours for
the blueshift and ionisation parameter of the absorber obtained with this analysis are
also shown in Figure \ref{fig_zabs} (dashed lines). Unsurprisingly, in this case
the fit cannot distinguish between the \fexxv\ and \fexxvi\ solutions without the
lower-energy feature to help determine the ionisation parameter. Consequently, the
range of allowed blueshifts is broader, but an extreme velocity is required regardless
of whether the line is associated with \fexxv\ or \fexxvi. Similar re-analysis of the
\nsims\ simulated datasets finds that only \ngtrexcl\ returned a chance improvement
equivalent to or greater than that observed at any velocity searched after excluding
this energy range, implying that the outflow is still detected at $\sim$98.5\%
significance regardless of the nature of the feature at $\sim$1.5\,keV; the
$\Delta\chi^{2}$ distribution obtained with this analysis is also shown in Figure
\ref{fig_sims}.

Second, we consider whether there are likely to be any systematic issues regarding
either our background subtraction or our modeling that could result in an artificial
feature at $\sim$1.5\,keV. To do so, we first varied the position of the background
region, and found both the structure in the background-subtracted spectrum at
$\sim$1.5\,keV and the $\Delta\chi^{2}$ improvement provided by the ionised absorber
to be insensitive to these variations. This is not surprising, as the spatial distrubition of
the background aluminium emission across the \xmm\ field of view is known to be
stable\footnote{http://xmm2.esac.esa.int/docs/documents/CAL-TN-0066-0-0.pdf}. The
same structure is also present if we adopt a more standard (and stricter) filtering of
periods of high background (instead of using the method outlined in \citealt{Picon04}).
The only systematic possibility that remains is that there is an error in the overall
background level, and so we also repeated our analysis with a smaller \xmm\
extraction region (for both \epicpn\ and \epicmos), reducing the radius from 25$''$ to
20$''$. While this lowers the source counts by $\sim$5--10\%, the background is
reduced by $\sim$35\%. Again, we find that the same structure is seen in the 1--2\,keV
band, and the statistical improvement provided by the addition of the ionised absorber
is practically the same as that reported above, $\Delta\chi^{2} = 35$. We also
find no difference in the improvement provided by the ionised absorber if we allow the
instrumental gain to vary for the \xmm\ datasets.

We also tested whether the presence of this additional absorption feature was
influenced by the parameters assumed for the \mekal\ component. As discussed
above, with the exception of iron (which is linked to the other model components
and is free to vary), in our modeling we assume solar abundances for the elements
included in the \mekal\ model. However, since iron is found to have a non-solar
abundance, it is also possible that this would also be the case for other elements,
and given that a number of the elements included in \mekal\ have lines close to
$\sim$1.5\,keV (\eg\ magnesium and silicon), incorrectly assuming a solar
abundance could potentially produce residuals consistent with an absorption feature
in the EPIC data. To be conservative, we therefore repeated our fits allowing the
abundances for all elements with atomic numbers between oxygen and calcium to
vary between 0.1--10.0 times the solar value. We find that none of these abundances
are well constrained (so we retain the assumption of solar abundances in the best-fit
models presented), the structure in the 1--2\,keV band is still seen, and that the
statistical improvement provided by the ionised absorber is again similar to that
reported above, $\Delta\chi^{2} = 32$.

Finally, we also use our simualtions to assess the likelihood that the feature at
$\sim$1.5\,keV could be produced by a statistical fluctuation of the \xmm\ aluminium
background emission (although we note that the same basic structure is seen in the
$\sim$1--2\,keV energy range in both the \epicpn\ and \epicmos\ data, which already
implies that a pure statistical fluctuation is unlikely; see Figure \ref{fig_lines}).
Modeling this feature with a Gaussian absorption line, we find an observed-frame
energy of $1.51 \pm 0.05$\,keV and that the line is not well resolved ($\sigma <
0.13$\,keV), with an equivalent width of $130^{+90}_{-60}$\,eV. The addition of this
line improves the fit by $\Delta\chi^{2} = 23$ in the real data (so this lower-energy
features make a slightly stronger statistical contribution to the total improvement
provided by the \xstar\ grid). For each of our \nsims\ simulations, we determine the
$\Delta\chi^{2}$ improvement provided by a Gaussian feature with a similar energy
and width (constrained to the ranges given above, since in this case we are testing
for the effects of a background line at a known energy), but allowing for the feature to
be in either absorption or emission (to reflect the fact that statistical fluctuations could
be either positive or negative). Again, we find that \ngtrAl\ of the simulated datasets
showed a chance improvement at or above the level seen in the real data; again, the
$\Delta\chi^{2}$ distribution is shown in Figure \ref{fig_sims}. This implies that there
is a $\lesssim$0.01\% chance that a statistical fluctuation of the background
aluminium line could produce a feature similar to that observed.

Based on these tests we therefore conclude that, although there is a background
feature at a similar energy and the spectral model at low energies is complex, there
is little evidence that the additional absorption feature seen in the source spectrum at
$\sim$1.5\,keV is purely the result of a poor background subtraction (either systematic
or statistical) or an artefact of our modeling. We are therefore confident that our
detection of a highly blueshifted, highly ionised absorber in \iras\ is robust.

\section{Discussion}
\label{sec_dis}

We have presented a spectral analysis of a deep, broadband X-ray observation
of the Seyfert 2 galaxy \iras, taken by \xmm\ and \nustar\ in coordination. Although
during this epoch the source was found to be significantly fainter than previous
observations (by a factor of $\sim$6), the data still reveal that the X-ray spectrum of
\iras\ is complex, showing contributions from relativistic disc reflection, absorption and
additional reprocessing by more distant material, and ionised absorption from an
ultrafast outflow. This complexity is analogous to the better-studied AGN in NGC\,1365
(\citealt{Risaliti13nat, Walton14, Rivers15}) and IRAS\,13197--1627
(\citealt{Miniutti07iras, Walton18}). The \nustar\ data presented here provide the first
high-energy ($E > 10$\,keV) detection of this source, which is critical for disentangling
the effects of these various processes.

\subsection{The Inner Accretion Disc}

From the relativistic reflection, we are able to place constraints on the key parameters
for the innermost accretion flow. With regards to the black hole spin, we find that a
rapidly rotating black hole is preferred ($a > 0.77$; Figure \ref{fig_spin}),
with tighter constraints if we assume a standard electron temperature of $kT_{\rm{e}}
= 40$\,keV for the corona ($a > 0.91$). This is a quantity of particular interest, as it
provides a window into the growth history of the central supermassive black hole
(\eg\ \citealt{Sesana14, Dubois14, Fiacconi18}). A high spin parameter implies the
black hole primarily grew through a major phase of coherent accretion (perhaps
triggered by a major merger or just through primarily feeding on gas in its host galaxy),
as opposed to having a more chaotic growth history (e.g. growing through a large
number of minor mergers). We also infer that we view the accretion disk at a
relatively high inclination ($i = 59^{+3}_{-16}$\deg), which is roughly consistent with
its Seyfert 2 classification in the standard unified model for active galaxies (\eg\
\citealt{AGNunimod}).

We stress that the spin constraint presented here is completely driven by the strength
of the reflected emission, which is inferred to be high ($R_{\rm{frac}} > 1.7$); if we
remove the functional connection between the spin parameter and the reflection
fraction, then we find the spin to be unconstrained. As such, our spin constraint is
heavily dependent on the assumed thin disc geometry, which requires strong
gravitational lightbending, and in turn a combination of a compact corona and a rapidly
rotating black hole, to produce a large reflection fraction (\citealt{lightbending,
Parker14mrk, Dauser14}). If the real geometry differs from this substantially, for
example if the disc has a large scale height (as may be expected at high accretion
rates relative to the Eddington limit; \citealt{Shakura73}), then our spin constraint
should be viewed with some caution. However, we note that the observation
presented here caught \iras\ in a low-flux state (a factor of $\sim$6 fainter than prior
X-ray osbervations). This does not appear to be related to large changes in the
line-of-sight absorption, and may therefore imply a lower accretion rate through the
inner disc than seen previously. Even if \iras\ was close to Eddington in its high-flux
observations, the thin disc approximation may therefore still be reasonable during
this epoch. Furthermore, the spin inferred here is consistent with the previous
estimates presented by \cite{Tan12} and \cite{Ricci14}, which also imply a rapidly
rotating black hole with $a > 0.73$ based solely on the profile of the iron emission;
we note that the conditions seen in \iras\ (large reflection, compact corona) are the
optimum conditions for obtaining spin constraints (\citealt{Bonson16, Kammoun18}).
Although the line profile can also be influenced by these same issues relating to the
assumed geometry, \cite{fenrir} show that assuming the disc is thin when in reality it
has some thickness tends to result in the spin being slightly underpredicted, so this
should not change the consistency of the results. Nevertheless, an independent mass
measurement for \iras\ is required to robustly determine its Eddington limit and thus
its true accretion regime.

\subsection{The Ultrafast Outflow}

The other main result of interest is that we find strong evidence for an extremely rapid,
ionised outflow. This imprints absorption features at $\sim$1.5 and $\sim$9.5\,keV
(observed frame) from ionised iron, which can be explained with a common blueshift
of $z_{\rm{abs}} = -0.349 \pm 0.009$ (relative to the cosmological redshift of \iras)
and an ionisation parameter of $\log[\xi/(\rm{erg~cm~s}^{-1})] = 4.7 \pm 0.1$ (see
Figure \ref{fig_zabs}). Applying the relativistic Doppler formula, we find a line-of-sight
velocity of  $\beta_{\rm{LoS}} = v_{\rm{LoS}}/c = -0.405 \pm 0.012$. If this is
associated with an outflow, this would be one of the most extreme outflows currently
known in terms of its observed velocity; only PDS\,456 is currently known to have a
more blueshifted component to its outflow ($\beta_{\rm{out}} = 0.46 \pm 0.02$;
\citealt{Reeves18}). Given the broadly equatorial geometry expected for winds from
an accretion disc (\eg\ \citealt{Proga04, Ponti12}), and the high inclination we infer
from the reflected emission, it is likely that the true outflow velocity is close to that
projected onto our line-of-sight, \ie $\beta_{\rm{out}} \sim \beta_{\rm{LoS}}$.

The kinetic luminosity of the outflow, $L_{\rm{kin}}$ ($= 1/2 \dot{M}v_{\rm{out}}^2$), is
given by Equation \ref{eqn_wind}, based on the standard expression for
$\dot{M}_{\rm{out}}$ for outflowing material derived by considering conservation
of mass arguments:

\begin{equation}
L_{\rm{kin}}\approx2{\pi}{\Omega}C_{\rm{V}}m_{\rm{p}}{\mu}nR^{2}v_{\rm{out}}^3
\label{eqn_wind}
\end{equation}
\vspace*{0.1cm}

\noindent{Here} $C_{\rm{V}}$ is the volume filling factor of the wind (a measure of its
`clumpiness'; normalised such that $0 \leq C_{\rm{V}} \leq 1$, similar to $\Omega$),
$m_{\rm{p}}$ is the proton mass, and $\mu$ is the mean atomic weight ($\sim$1.2 for
solar abundances; given the super-solar iron abundance inferred $\mu$ may in reality
be a little higher in this case).

The key question regarding AGN outflows is whether they carry sufficient power to
drive the feedback invoked to explain known correlations between AGN and their
host galaxies. This is usually determined by estimating the ratio between the kinetic
power and the bolometric radiative luminosity. However, several of the quantities in
Equation \ref{eqn_wind} are notoriously difficult to estimate: $n$, $R$, $\Omega$ and
$C_{\rm{V}}$. In rare cases, it is possible to constrain some of these parameters
directly (\eg\ PDS\,456; \citealt{Nardini15}), but typically we are forced to re-frame
Equation \ref{eqn_wind} in terms of quantities that are more readily observable. We
can combine this with the defnition of the ionisation parameter to write an expression
for $L_{\rm{kin}}/L_{\rm{bol}}$ in terms of $\xi$ (similar to, \eg, \citealt{Pinto16nat,
Walton16ufo, Kosec18}):

\begin{equation}
\frac{L_{\rm{kin}}}{L_{\rm{bol}}}\approx2{\pi}m_{\rm{p}}\mu\frac{v_{\rm{out}}^3 L_{\rm{ion}}}{{\xi}L_{\rm{bol}}}{\Omega}C_{\rm{V}}
\label{eqn_wind_ion}
\end{equation}
\vspace*{0.1cm}

Given the intrinsic SED adopted for \iras\ (see Section \ref{sec_ufo}, we assume that
the ratio $L_{\rm{ion}}/L_{\rm{bol}} \sim 1$. From the constraints on $v_{\rm{out}}$
and $\xi$, we therefore estimate that $L_{\rm{kin}}/L_{\rm{bol}} \sim 500 \Omega
C_{\rm{V}}$. Although we do not know either $\Omega$ or $C_{\rm{V}}$ here, the
above $L_{\rm{kin}}/L_{\rm{bol}}$ estimate is comparable to similar calculations for
the winds being seen in ultraluminous X-ray sources (ULXs; \eg\ \citealt{Pinto16nat,
Pinto17, Walton16ufo, Kosec18}). These are now generally accepted to be
high/super-Eddington accretors (\eg\ \citealt{Pintore17, Koliopanos17,
Walton18ulxBB, Walton18p13}), particularly after the discovery that some of these
sources are powered by neutron stars (\citealt{Bachetti14nat, Fuerst16p13, Israel17,
Israel17p13, Carpano18}). This may further support the conclusions of \cite{Ricci14}
that, at least at its brightest, \iras\ is a high-Eddington source, around which we have
based a number of our calculations. 

This approach for estimating $L_{\rm{kin}}/L_{\rm{bol}}$ essentially assumes that all
of the absorbing material is located at a single radius (\ie $\Delta R \ll R$) that satisfies
the definition of $\xi$, and should likely be considered an upper limit. Alternatively, we
can use the fact that the column density is a line-of-sight integration of the density to
write another expression for $L_{\rm{kin}}/L_{\rm{bol}}$ in terms of $N_{\rm{H}}$
adopting the opposite limiting geometry, \ie $\Delta R \sim R$ (similar to, \eg,
\citealt{Krongold07, Crenshaw12, Nardini15}):

\begin{equation}
\frac{L_{\rm{kin}}}{L_{\rm{bol}}}\approx2{\pi}m_{\rm{p}}\mu\frac{RN_{\rm{H}}v_{\rm{out}}^3}{L_{\rm{bol}}}{\Omega}C_{\rm{V}}
\label{eqn_wind_nh}
\end{equation}
\vspace*{0.1cm}

\noindent{Here} we are still left with a factor of $R$, which is also not known. However,
we can place a conservative lower limit on $R$ by equating the outflow velocity to the
escape velocity (similar to \citealt{Nardini18}). For $v_{\rm{out}} = 0.4c$, we
find that $R > 12.3$\,\rg, or $R \gtrsim 9 \times 10^{11}$\,m for our estimated black
hole mass of $M_{\rm{BH}} \sim 5 \times 10^{7}$ \msun\ (Section \ref{sec_ufo}). From
the constraints on $v_{\rm{out}}$ and $N_{\rm{H}}$, and taking $L_{\rm{bol}} \sim
10^{45}$\,\ergps\ for this epoch (based on the bolometric correction discussed in
Section \ref{sec_ufo}), we therefore estimate that $L_{\rm{kin}}/L_{\rm{bol}} \gtrsim 0.1
\Omega C_{\rm{V}}$.

Although there are clearly still large uncertainties related to the geometry of the wind,
unless the product $\Omega C_{\rm{V}}$ is small the ultra-fast outflow in \iras\ should
easily be sufficient to drive AGN feedback on galactic scales (which requires
$L_{\rm{kin}}$ to be greater than $\sim$a few \% of $L_{\rm{bol}}$;
\citealt{DiMatteo05nat, Hopkins10}), and may even dominate the total energy output
from the system. We note that for PDS\,456, the solid angle of the ultra-fast outflow is
estimated to be $\Omega \sim 0.75$ (\citealt{Nardini15}). Although the volume filling
factor is still not formally known in that case, this must presumably also be large since
the wind is persistently observed (\citealt{Matzeu17}). Assuming \iras\ is close to is
Eddington limit, as expected for PDS 456, it is plausible that its winds would be similar,
also with a large solid angle. \cite{Nardini15} also estimated the radius of the wind in
PDS\,456 to be $\sim$100\,\rg. Repeating the calculation above and assuming this
radius and solid angle for \iras, we infer $L_{\rm{kin}}/L_{\rm{bol}} \sim 0.6 C_{\rm{V}}$.


The most extreme component of the PDS\,456 outflow has only been seen when the
source was in a low-flux state (\citealt{Reeves18}). This is also similar to the ultrafast
outflow seen in the NLS1 IRAS\,13224--3809, which produces significantly stronger
absorption features when the flux is low (\citealt{Parker17nat, Pinto18iras,
Jiang18iras}), potentially related to the ionisation of the wind responding to changes
in the source flux. Although we cannot currently address whether the strong outflow
reported here would also be observable when \iras\ has a high flux, owing to a
combination of S/N and bandpass issues with the available high-flux data, it is
interesting to note that this has also been observed while the source was in a low-flux
state, potentially similar to those better-studied cases (which are also high-Eddington
accretors). Future broadband observations of \iras\ at higher fluxes will be needed to
address this, and shed further light on the conditions in and/or the geometry of the
wind.

Finally, we note that an alternative explanation for these highly blueshifted absorption
features that does not require any kind of outflow has been proposed by \cite{Gallo11,
Gallo13abs}, who suggest that they may arise through absorption in clouds suspended
(potentially magnetically) above the disc, and co-rotating with it. This idea has recently
been further explored by Fabian et al. (2018; submitted), and is primarily relevant for
discs extending to the ISCO of a rapidly rotating black hole with very compact coronae
(such that the illumination of the disc is strongly centrally concentrated and the
contribution of the reflection to the observed spectrum is strong) and viewed at
high inclination, broadly similar to the scenario inferred here. Such a configuration
results in an apparent shift in the energy of the absorption line (rather than a
broadening of it) as the reflected emission observed from the disc is almost
completely dominated by the blue side, and so we only see absorption from material 
along this line-of-sight. Although velocities of up to 0.5$c$ are naturally present within
the disc, the requirement that the absorbing medium be along our line-of-sight to the
point of maximum emission likely sets an upper limit to the velocity shifts this model
can reasonably produce that is lower than this. The velocity shift of $\beta_{\rm{LoS}}
\sim 0.4$ inferred here is extreme and would likely push this interpretation to its limits,
particularly given that the inclination is only moderately high, so an interpretation
invoking an outflow (as discussed above) is likely preferred in this case.


\section{Conclusions}
\label{sec_conc}

The deep \xmm+\nustar\ observation of \iras\ taken in 2017 shows evidence for
relativistic reflection from the inner accretion disc, neutral absorption, further
reprocessing by more distant material, and ionised absorption in an extreme,
ultrafast outflow ($v_{\rm{out}} \sim 0.4c$). By modeling the disc reflection with
simple lamppost models, we find that the central supermassive black hole is likely
rapidly rotating ($a > 0.77$), consistent with previous estimates from the profile of
the relavitistic iron line. We also find that the accretion disc is viewed at a fairly high
inclination ($i \sim 59$\deg). The energetics of the ultrafast outflow are still highly
uncertain, but we estimate that it is likely sufficient to power galaxy-scale AGN
feedback, and may even dominate the total energetics of the system.

\section*{ACKNOWLEDGEMENTS}

The authors would like to thank the reviewer for their feedback, which helped 
improve the clarity of the final version of the manuscript, and C. Done for useful
discussion.
DJW acknowledges support from an STFC Ernest Rutherford Fellowship; EN
acknowledges funding from the European Union's Horizon 2020 research and
innovation programme under the Marie Sk\l{}odowska-Curie grant agreement no.
664931; CR acknowledges the CONICYT+PAI Convocatoria Nacional subvencion a
instalacion en la academia convocatoria a\~{n}o 2017 PAI77170080; ACF
acknowledges support from ERC Advanced Grant 340442; JAG acknowledges
support from NASA grant NNX17AJ65G and from the Alexander von Humboldt
Foundation. 

This research has made use of data obtained with \nustar, a project led by Caltech,
funded by NASA and managed by NASA/JPL, and has utilized the \nustardas\
software package, jointly developed by the ASDC (Italy) and Caltech (USA). This
research has also made use of data obtained with \xmm, an ESA science mission
with instruments and contributions directly funded by ESA Member States.


\bibliographystyle{/Users/dwalton/papers/mnras}

\bibliography{/Users/dwalton/papers/references}

\label{lastpage}

\end{document}